\documentclass[aps,pra,showpacs,twocolumn,superscriptaddress]{revtex4-1}
\usepackage{bm,color,amsmath,amssymb,mathrsfs,latexsym,graphicx,psfrag}
\usepackage[utf8]{inputenc}
\usepackage{bbm}
\usepackage[normalem]{ulem}

 \newcommand{\imag}{\mathrm{i}}

\begin{document}
	\title{ $SU(3)$ Topological Insulators in the Honeycomb Lattice}
	\author{U. Bornheimer}
	\affiliation{Centre for Quantum Technologies, National University of Singapore, 3 Science Drive 2, 117543 Singapore}
	\affiliation{MajuLab, CNRS-UNS-NUS-NTU International Joint Research Unit, UMI 3654, Singapore}
	\affiliation{Department of Physics, Faculty of Science, National University of Singapore, 2 Science Drive 3, 117551 Singapore}

	\author{C. Miniatura}
	\affiliation{MajuLab, CNRS-UNS-NUS-NTU International Joint Research Unit, UMI 3654, Singapore}
	\affiliation{Centre for Quantum Technologies, National University of Singapore, 3 Science Drive 2, 117543 Singapore}
	\affiliation{Department of Physics, Faculty of Science, National University of Singapore, 2 Science Drive 3, 117551 Singapore}
	\affiliation{School of Physical and Mathematical Sciences, Nanyang Technological University,
Singapore 637371, Singapore}
	\affiliation{Université Côte d’Azur, CNRS, InPhyNi; France}
	\affiliation{Institute of Advanced Studies, Nanyang Technological University, 60 Nanyang View, Singapore 639673, Singapore}

	\author{B. Grémaud}
	\affiliation{MajuLab, CNRS-UNS-NUS-NTU International Joint Research Unit, UMI 3654, Singapore}
	\affiliation{Centre for Quantum Technologies, National University of Singapore, 3 Science Drive 2, 117543 Singapore}
	\affiliation{Department of Physics, Faculty of Science, National University of Singapore, 2 Science Drive 3, 117551 Singapore}
	\affiliation{Laboratoire Kastler-Brossel, Ecole Normale Supérieure CNRS, UPMC; 4 Place Jussieu, F-75005 Paris, France}

\begin{abstract}
	We investigate realizations of topological insulators
	with spin-1 bosons loaded in a  honeycomb optical 
 lattice and subjected to a $SU(3)$ spin-orbit coupling - a situation which 
 can be realized experimentally using cold atomic gases. 
     In this paper, we focus on the topological properties of
	the single-particle band structure, namely Chern numbers 
	(lattice with periodic boundary conditions) and edge states (lattice with strip geometry). 
	While $SU(2)$ spin-orbit couplings always lead to time-reversal symmetric Hubbard models, 
	and thereby to topologically trivial band structures, suitable $SU(3)$ spin-orbit 
	couplings can break time reversal symmetry and lead to topologically non-trivial bulk band structures and 
	to edge states in the strip geometry. In addition, we show that one can trigger a series of 
	topological transitions (i.e. integer changes of the Chern numbers) that are specific 
	to the geometry of the honeycomb lattice by varying a single parameter in the Hamiltonian.
\end{abstract}

\maketitle

\section{Introduction}

Over the last few years, the continuous progress in the degree of control, tunability and flexibility of 
ultracold atomic gases experiments \cite{Lewenstein07,Blochreview08,Ketterle2} has opened the lab door to 
a whole class of model Hamiltonians, as witnessed for example by the recent implementation of artificial gauge  
fields~\cite{Spielman09a,Spielman09b,Spielman11,Zhang12,Zwierlein12,Windpassinger12,Spielman12}. 
Some of these model Hamiltonians are directly inherited from 
condensed matter physics, for instance the integer and fractional quantum Hall
effects~\cite{FCI13,FCIPRL1,FCIPRL2}. More saliently, physicists have further proposed new theoretical 
ideas and physical situations, such as topological phases~\cite{Mottonen09,Bercioux11,Hueda12,YuXin15,Goldman2016,Zhai2015}, 
non-Abelian particles~\cite{Burrello10} or  
mixed dimensional systems~\cite{Nishida_2008,Nishida_2010,Lamporesi_2010,Huang_2013,Iskin_2010}, that 
could be tested in the lab. 
In particular, experiments involving spinors, either made of bosons or fermions in different Zeeman sub-levels,
are now able to implement and study non-Abelian gauge fields~\cite{Goldman2016,Livi2016,Gross2017,Sugawa2016,Ray2014}. 
In these systems, the kinetic 
energy term in the Hamiltonian  allows for a modification of the internal degrees of freedom as the 
particle propagates~\cite{Dalibard11,Goldman2013},  leading to a rich class of non trivial physical 
phenomena, especially with interactions. 
In two-dimensional lattices, and in particular when spin-orbit coupling is present, the corresponding 
non-Abelian gauge fields induce non-diagonal hopping matrices in the tight-binding Hubbard Hamiltonian 
that mix and flip the spin degrees of freedom~\cite{Dalibard11,Goldman2013}.
2-component bosonic and fermionic gases (in the bulk or in a lattice) have been the subject of many 
recent analytical and numerical 
studies~\cite{Lewenstein09,Cole_2012,Cai12,Galitski12,Hofstetter12,Fujimoto09,Goldman12,Jiri2014,
Shenoy11,Sademelo12,Lewenstein10,Iskin11,Iskin12,Iskin13,Riedl13}.

In marked contrast, 3-component bosonic or fermionic gases subjected to a $SU(3)$ gauge field have 
been much less studied: the experimental
realization is more complicated~\cite{Dalibard11,YuXin2014,Kosior2014} and their theoretical studies
are more involved, the gauge field group being much larger~\cite{Barnett2012,Mandal2016,Han2016,Grass2014,Go2013}.
On the other hand, 
tight-binding models with a $SU(3)$ spin-orbit coupling can break time reversal symmetry and lead to topological 
insulators: the bulk band structure is topologically non trivial (non-zero Chern numbers) and edge states 
develop for a strip geometry~\cite{Barnett2012}. Such a situation cannot occur in any kind 
of $SU(2)$ models as we will explain later. In addition, since $SU(3)$ has a more complex group 
structure than $SU(2)$, one expects a larger
variety of spin textures for interacting particles. These spin textures are associated to different 
homotopy groups and appear both in the ground state and in the excitations above the ground 
state~\cite{Hueda12,Ezawa_book,Mandal2016,Han2016,Grass2014,Go2013}. 

Our paper consists of three main parts. In Section~\ref{sec:method}, we give some general 
properties of $SU(N)$ Hubbard models and describe the $SU(3)$ tight-binding model on the 
honeycomb lattice that we consider. We analyze its time-reversal properties. In Section~\ref{sec:topo} 
we study the topological properties of the band structure obtained for a lattice with periodic 
boundary conditions and compute the corresponding Chern numbers~\cite{Resta2000, Paly2016, Niu2010, Houches2014}. 
Next we study the edge states that are expected in the lattice strip geometry, as inferred 
from the bulk-edge correspondence~\cite{Hatsugai1993a, Hatsugai1993b, Hatsugai1997, Kane2005, Kane2010}. 
In Section~\ref{sec:transition}, we show how to trigger a series of topological transitions in the 
band structure of our $SU(3)$ model (i.e. integer changes in the Chern numbers) by varying a single 
parameter in the Hamiltonian. The variety of such topological transitions is richer on the honeycomb 
lattice than on the square lattice. In Section~\ref{sec:conclusion}, we summarize our results and 
conclude with some perspectives.

\section{$SU(N)$ tight-binding models and time-reversal symmetry}
\label{sec:method}

\subsection{Some general properties}

The general tight-binding Hubbard Hamiltonian for non-interacting particles with spin $s$ on a 
two-dimensional lattice has the following form

\begin{align}
	\hat{H}= \sum_{\langle i,j\rangle} \hat{\bm{\psi}}_{i}^{\dagger}T_{ij}\hat{\bm{\psi}}_{j},
	\label{eq:hamil}
\end{align}
where the sum is carried over all nearest-neighbour lattice site pairs $\langle i,j\rangle$ and where 
${\hat{\bm{\psi}}_i^{\dagger}=\left(\hat{\psi}_{i,s}^{\dagger},\hdots,\hat{\psi}_{i,-s}^{\dagger}\right)}$ 
is the $(2s+1)$-component row-spinor built at each lattice site $i$ on 
the creation operators $\hat{\psi}^{\dagger}_{i,\sigma}$ for each spin 
component $\sigma$ ($|\sigma| \leq s$). The $(2s+1)\times(2s+1)$ hopping matrix $T_{ij}$ is 
connecting the different spin components at site $j$ to the different spin components at site $i$. 
Since the Hamiltonian $\hat{H}$ is Hermitian, one has $T_{ji}=T^{\dagger}_{ij}$. $SU(N)$ 
Hubbard models ($N=2s+1$) are obtained with $T_{ij}=-t_{ij} U_{ij}$ where $t_{ij}$ are real 
positive numbers and where the matrices $U_{ij} \in SU(N)$ describe the unitary transformation of 
the spin states between sites $j$ and $i$. Using the exponentiation map, 
one often writes $U_{ij} = \exp{(\imag A_{ij})}$ where $A_{ij}$ are $N\times N$ 
traceless Hermitian matrices representing the gauge field acting on the system. 
The matrices $A_{ij}$ live in the Lie algebra spanned by the generators of 
$SU(N)$~\cite{GilmoreBook}. Since $T_{ji}=T^{\dagger}_{ij}$, 
we have $t_{ji} = t_{ij}$ and $U_{ji}=U^{\dagger}_{ij}$, which implies $A_{ji} = -A_{ij}$. 
Non-Abelian $SU(N)$ models are obtained when the matrices $U_{ij}$ do not commute, i.e. 
when the matrices $A_{ij}$ do not commute, meaning that a non-Abelian gauge field is acting on the system.

When the system is invariant under a certain discrete translation group $\mathcal{T}_{{\bf r}}$, 
the spectrum of the Hamiltonian $\hat{H}$ exhibits a band structure. Note that $\mathcal{T}_{{\bf r}}$ can differ from
the Bravais translation group of the underlying lattice itself (this is the case, 
for example, in the presence of an external magnetic field with rational flux per plaquette).
Each band $n$ is described by its eigenvalues 
$\epsilon_n(\mathbf{k})$ and eigenvectors $|n,\mathbf{k}\rangle$ for all Bloch wavevectors $\mathbf{k}$ 
in the first Brillouin zone $BZ$ defined by the translation group $\mathcal{T}_{{\bf r}}$. 
For an isolated band, one can define the first Chern number as~\cite{Niu2010, Houches2014}
\begin{align}
	C_n=&\frac{1}{2\pi}\int_{BZ}\mathrm{d}^2\mathbf{k}\ \Omega_n(\mathbf{k}),\label{eq:chern}
\end{align}
where $\Omega_n(\mathbf{k})$ is the Berry curvature of the $n$th band 
\begin{align}
	\Omega_n(\mathbf{k})=\left(\nabla_{\mathbf{k}}\times\underbrace{i\langle n,\mathbf{k}|
	\nabla_{\mathbf{k}}|n,\mathbf{k}\rangle}_{\mathbf{A_n}(\mathbf{k})}\ \right)\cdot\mathbf{e}_z,
\end{align}
and $\mathbf{A}_n$ is the Berry connection of the $n$th band~\cite{Niu2010}. 
Being an integer, the Chern number can only change when two bands touch, that is, when a gap closes.
Chern numbers of a given tight-binding Hamiltonian satisfy a "zero-sum rule": 
they all add up to zero. 
For inversion-symmetric systems (resp. for time-reversal invariant systems), 
it is well known that the Berry curvature is even (resp. odd) in the $BZ$:
\begin{align*}
		\Omega_n(\mathbf{k})=\left\lbrace\begin{matrix}
			+\Omega_n(-\mathbf{k})\ \mathrm{inversion\ symmetry}\phantom{blah}\\
			-\Omega_n(-\mathbf{k})\ \mathrm{time\ reversal\ symmetry}
		\end{matrix}\right..
\end{align*}
These two properties show that: (i) the Berry curvature itself identically vanishes if $\hat{H}$ 
is invariant under both space inversion and time reversal; (ii) Chern numbers of time-reversal invariant Hamiltonians 
are necessarily vanishing. Therefore, only systems that break time reversal symmetry can 
have non-trivial topological properties,
i.e. non-vanishing Chern numbers. In the present situation, the topological properties of a tight-binding Hamiltonian
$\hat{H}$, see Eq.~\eqref{eq:hamil}, depend 
crucially on the behavior of the different $T_{ij}$ under time reversal.

At the single particle level, and up to an inessential overall phase factor, 
the (anti-unitary) time reversal operator is defined as 
${\Theta = e^{-\frac{\imag\pi}{\hbar}J_y}K}$, where $J_y$ is the 
projection of the spin operator ${\bf J}$ of the particle on the $y$ direction and where $K$ is
the complex conjugation operator. One has $\Theta^2=\openone$ for integer spins and $\Theta^2= - \openone$ for
half-integer spins~\cite{Sakurai1986}.  Under time reversal a $SU(N)$ Hamiltonian becomes~\cite{Bernevig2013} 
\begin{align*}
	\Theta \hat{H} \Theta^{-1}= - \sum_{\langle i,j\rangle} t_{ij} 
	\hat{\bm{\psi}}_{i}^{\dagger}e^{-\imag \Theta A_{ij} \Theta^{-1}} \hat{\bm{\psi}}_{j}  
\end{align*} 
and time reversal $\Theta \hat{H} \Theta^{-1}= \hat{H}$ 
is automatically guaranteed when the matrices $A_{ij}$ are all 
{\it odd} under time reversal. In this case, the corresponding $SU(N)$ Hubbard model is topologically trivial. 

This is the situation for spin-$\frac{1}{2}$ systems, where all the generators of SU(2) (Pauli-matrices) 
are odd under time reversal symmetry. In this case $\Theta=-\imag \sigma_yK$ where $\sigma_y$ 
is the projection of the Pauli spin operator $\boldsymbol{\sigma}$ along axis $y$. 
As a result, {\it all} non-interacting $SU(2)$ Hubbard models are topologically trivial. 
Indeed, for $SU(2)$ systems, $A_{ij} = \bm{\alpha}_{ij} \cdot \boldsymbol{\sigma}$ where $\bm{\alpha}_{ij}$ 
are real vectors and $A_{ij} $ is always odd under 
time reversal since $\Theta \boldsymbol{\sigma} \Theta^{-1}= - \boldsymbol{\sigma}$.

On the contrary, for $N\geq 3$, that is for spins $s \geq 1$, the even sector of the Lie algebra $SU(N)$, 
made by the operators $\mathcal{O}$ that fulfill $\Theta \mathcal{O} \Theta^{-1}= \mathcal{O}$,
is not empty. Thus, by conveniently choosing $A_{ij}$ in the even sector, 
one can break time reversal invariance and have topologically non-trivial $SU(N)$ Hamiltonians. 
In this paper we will consider such a system for spin-$1$ particles. 
In this case, the time reversal operator is $\Theta = J K$ with: 
\begin{align}
	J = \left(\begin{matrix}
		0&0&1\\
		0&-1&0\\
		1&0&0
	\end{matrix}\right).
\end{align}
Using the generators of the $SU(3)$ group, one can write 
$A_{ij} = \bm{\alpha}_{ij} \cdot \boldsymbol{\lambda}$, 
where $\bm{\alpha}_{ij}$ is a $8$-component real vector and 
$\boldsymbol{\lambda}$ the Gell-Mann vector made of the $8$ Gell-Mann matrices $\lambda_a$ 
(see Appendix~\ref{app:GMM}). 
From the transformation properties of the Gell-Mann matrices under time reversal, 
one finds that the odd sector of the Lie algebra is spanned by 
$(\lambda_1+\lambda_6)$, $(\lambda_2+\lambda_7)$ and $(\lambda_3+\sqrt{3} \lambda_8)$, 
while the even sector is spanned  by $\lambda_4$, $\lambda_5$, $(\lambda_1-\lambda_6)$, 
$(\lambda_2-\lambda_7)$ and $(\lambda_8-\sqrt{3} \lambda_3)$. 
Thus, by choosing suitable components of $\bm{\alpha}_{ij}$ in the even sector, 
one can produce gauge fields $A_{ij}$ breaking time reversal symmetry and leading to 
topologically non trivial $SU(3)$ Hubbard models, that is to Hamiltonians with a band structure 
having non-trivial Chern numbers~\cite{Barnett2012}. 
Note however that, while the time reversal invariance properties of the system can be directly and 
safely inferred from the matrices $U_{ij}$, it is a bit more subtle when using the exponential map. 
Indeed one could have $A_{ij}$ in the even sector and 
still preserve time reversal invariance provided that $\exp(2\imag A_{ij}) = \openone$. 
For instance  the gauge field 
$A_{ij} = \frac{\pi}{\sqrt{5}} \left[(\lambda_1-\lambda_6)+(\lambda_2-\lambda_7)+\lambda_4\right]$ leads to 
\begin{equation}
 U_{ij} = \exp(\imag A_{ij})=\frac{1}{5}\left(
 \begin{array}{ccc}
  -1 & 2+2\imag & -4\imag\\
  2-2\imag & -3 & -2-2\imag \\
  4\imag & -2+2\imag & -1 
 \end{array}\right),
\end{equation}
and one has $\Theta U_{ij}\Theta^{-1} = U_{ij}$.

In Appendix~\ref{app:TRU}, we give convenient parametrizations of unitary $U(3)$ 
matrices which are even or odd under time reversal.

\subsection{Non-Abelian $SU(3)$ model on the honeycomb lattice}

\begin{figure}[t!p]
	\includegraphics[width=\linewidth]{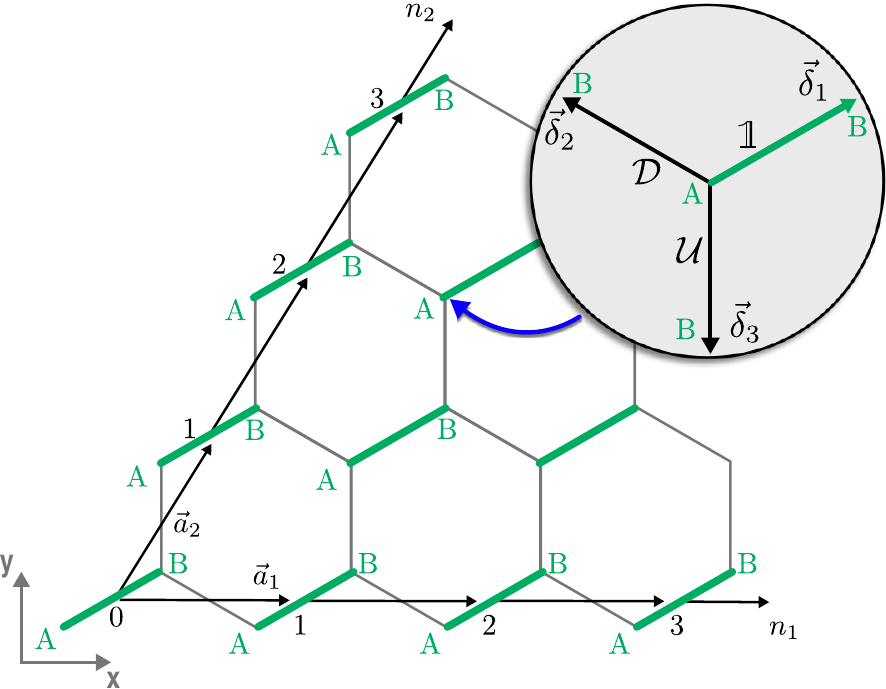}
	\caption{\label{fig:model}The non-Abelian $SU(3)$ model on the honeycomb lattice that we investigate. 
	The honeycomb lattice is obtained by repeated translations along Bravais vectors $\mathbf{a}_1$ and $\mathbf{a}_2$ 
	of a unit cell containing two inequivalent sites labeled $A$ and $B$ .
	We assign different spin-orbit couplings on the different $A$-$B$ links of the honeycomb lattice. 
	The inset shows the nearest-neighbor $SU(3)$ hopping matrices acting on the spin states of the particles: 
	$\openone$ along link vectors $\bm{\delta}_1$, diagonal hopping phase matrix $\mathcal{D}$ 
	along link vectors $\bm{\delta}_2$ and non-diagonal hopping matrix $\mathcal{U}$ along 
	link vectors $\bm{\delta}_3$, see main text for their expressions. }
\end{figure}

In the following, we investigate a $SU(3)$ topological insulator consisting of non-interacting  
spin-$1$ bosonic particles moving on a honeycomb lattice. Our system is similar to the one studied in 
Ref.~\cite{Barnett2012} on the square lattice. The honeycomb lattice is a triangular Bravais lattice 
obtained by repeated translations $\mathbf{R}(n_1,n_2)=n_1\mathbf{a}_1+n_2\mathbf{a}_2$ ($n_1$ and $n_2$ integers) 
of a unit cell containing two inequivalent sites denoted by $A$ and $B$, see Fig.~\ref{fig:model}. 
Of importance for the following are the link vectors $\bm{\delta}_1$, $\bm{\delta}_2$ and $\bm{\delta}_3$ 
connecting any site $A$ to its 3 nearest-neighbor sites $B$. They satisfy $\bm{\delta}_1-\bm{\delta}_2 = 
\mathbf{a}_1$, $\bm{\delta}_1-\bm{\delta}_3 = \mathbf{a}_2$ and $\bm{\delta}_1 + \bm{\delta}_2 +\bm{\delta}_3=0$, 
see Fig.~\ref{fig:model}. With our choice of origin in Fig.~\ref{fig:model}, 
the positions of all sites $A$ and $B$ in the lattice are labeled by 
$\mathbf{R}^A_{n_1,n_2} = \mathbf{R}(n_1,n_2)-\bm{\delta}_1/2$ and 
$\mathbf{R}^B_{n_1,n_2} = \mathbf{R}(n_1,n_2) + \bm{\delta}_1/2$. 
In the following we fix the unit length by setting the lattice constant to unity, i.e. $|\bm{\delta}_a|=1$ ($a=1,2,3)$.

We now assign the hopping matrices $T_1=-t \,\openone$ along links $\bm{\delta}_1$, $T_2=-t \,\mathcal{D}$ 
along links $\bm{\delta}_2$ and 
$T_3=-t \, \mathcal{U}$ along links $\bm{\delta}_3$, with 
$t$ a (real) hopping rate. 
The $SU(3)$ matrices $\mathcal{D}=\exp{(-\frac{2\pi \imag}{3}\hat{S}_z)}$ and 
$\mathcal{U}=\exp{\left[-\frac{2\pi \imag}{3\sqrt{3}}\left(\lambda_2-\lambda_5+\lambda_7\right)\right]}$ read
\begin{align}
\mathcal{D}=  \left(
					\begin{matrix}
						j^* & 0 & 0\\
						0 & 1 & 0\\
						0 & 0 & j
					\end{matrix}
				\right) \quad \quad
	\mathcal{U}=\left(
					\begin{matrix}
						0 & 0 & 1\\
						1 & 0 & 0\\
						0 & 1 & 0
					\end{matrix}
				\right)
\end{align}
where $j=\exp(2\imag \pi/3)$. From these expressions, one can check that $T_3$  is breaking time
reversal symmetry, which gives rise to a band structure with non-vanishing Chern number, see below Sec~\ref{sec:topo}. 
While spin states are unaffected when particles hop along link vectors $\bm{\delta}_1$, 
they acquire spin-state dependent phases when particles hop along link vectors $\bm{\delta}_2$ 
and they undergo a circular permutation $1 \to 0 \to -1 \to 1$ when particles hop along the link vectors $\bm{\delta}_3$.
Note that, since these matrices fulfill $\mathcal{D}^3 = \mathcal{U}^3 = \openone$, a spin state is mapped back
to itself after three consecutive hoppings along a given link vector $\bm{\delta}_2$ or $\bm{\delta}_3$.
On the contrary, since $\mathcal{D}$ and $\mathcal{U}$ do not commute,  the present spin-orbit coupling configuration 
corresponds to a genuine non-Abelian $SU(3)$ model on the honeycomb lattice: the
corresponding gauge fields $A_2$ and $A_3$ defined by $T_a=-t e^{\imag A_a}$ ($a = 2,3$) do not commute and,
therefore, the transport of a spin state around a hexagon leads to a non trivial Wilson loop value~\cite{Bermudez2010}.

\section{Topological properties}
\label{sec:topo}
\subsection{Infinite system - Bulk spectrum}

Since the unit cell of the honeycomb lattice hosts two inequivalent sites, it is customary
to distinguish the bosonic annihilation and creation operators on $A$ and $B$ sites: 
$\hat{\mathbf{a}}^{\dagger}_{n_1,n_2}$ denotes the creation operator on the site $\mathbf{R}^A_{n_1,n_2}$ and 
$\hat{\mathbf{b}}^{\dagger}_{n_1,n_2}$ denotes the creation operator on the site $\mathbf{R}^B_{n_1,n_2}$.
Each of them is a spinor of dimension 3 accounting for the 3 spin states of the 
spin-1 bosonic particles considered in our model. 

Being translation-invariant along the Bravais vectors $\mathbf{a}_1$ and $\mathbf{a}_2$, the lattice Hamiltonian is diagonal in momentum space, $\hat{H} = \sum_{\mathbf{k} \in BZ} \hat{H}_{\mathbf{k}}$, where
\begin{align}
\label{eq:bloch} 
\hat{H}_{\mathbf{k}} &= -t \, \left(\hat{\mathbf{b}}^{\dagger}_{\mathbf{k}} M_{\mathbf{k}} \hat{\mathbf{a}}_{\mathbf{k}}+ \mathrm{H.c.}\right) \\
M_{\mathbf{k}} &= e^{\imag\mathbf{k}\cdot\bm{\delta}_1} \, \openone + e^{\imag\mathbf{k}\cdot\bm{\delta}_2} \, \mathcal{D}
	+ e^{\imag\mathbf{k}\cdot\bm{\delta}_3} \, \mathcal{U}.
\end{align}
$\hat{\mathbf{a}}_\mathbf{k}$ (resp. $\hat{\mathbf{b}}_\mathbf{k}$) is the Fourier transform of 
$\hat{\mathbf{a}}_{n_1,n_2}$ (resp. $\hat{\mathbf{b}}_{n_1,n_2})$; the Bloch wavevector $\mathbf{k}$ belongs to 
the first Brillouin zone and reads
\begin{align}
	\mathbf{k}=\underbrace{k_1\mathbf{b}_1}_{\mathbf{k}_1}+\underbrace{k_2\mathbf{b}_2}_{\mathbf{k}_2},\ 
	|k_{1,2}|\ \leq 1/2,
\end{align}
where the honeycomb reciprocal lattice vectors $\mathbf{b}_1$ and $\mathbf{b}_2$ are 
defined by $\mathbf{a}_i\cdot\mathbf{b}_j=2\pi\delta_{ij}$.

Further defining $\hat{\Phi}_{\mathbf{k}}^{\dagger}=(\hat{\mathbf{a}}^{\dagger}_{\mathbf{k}}, \hat{\mathbf{b}}^{\dagger}_{\mathbf{k}})$, one can recast this Hamiltonian under the form $\hat{H}_{\mathbf{k}} = \hat{\Phi}_{\mathbf{k}}^{\dagger} \mathcal{H}_{\mathbf{k}} \hat{\Phi}_{\mathbf{k}}$ where 
\begin{align}
\label{Hamil}
\mathcal{H}_{\mathbf{k}}= - t \left(
					\begin{matrix}
						0 & M_{\mathbf{k}}^{\dagger}\\
						M_{\mathbf{k}} &  0
					\end{matrix}.
				\right)
\end{align}

\begin{figure*}[t!!!]
	\includegraphics[width=0.95\textwidth]{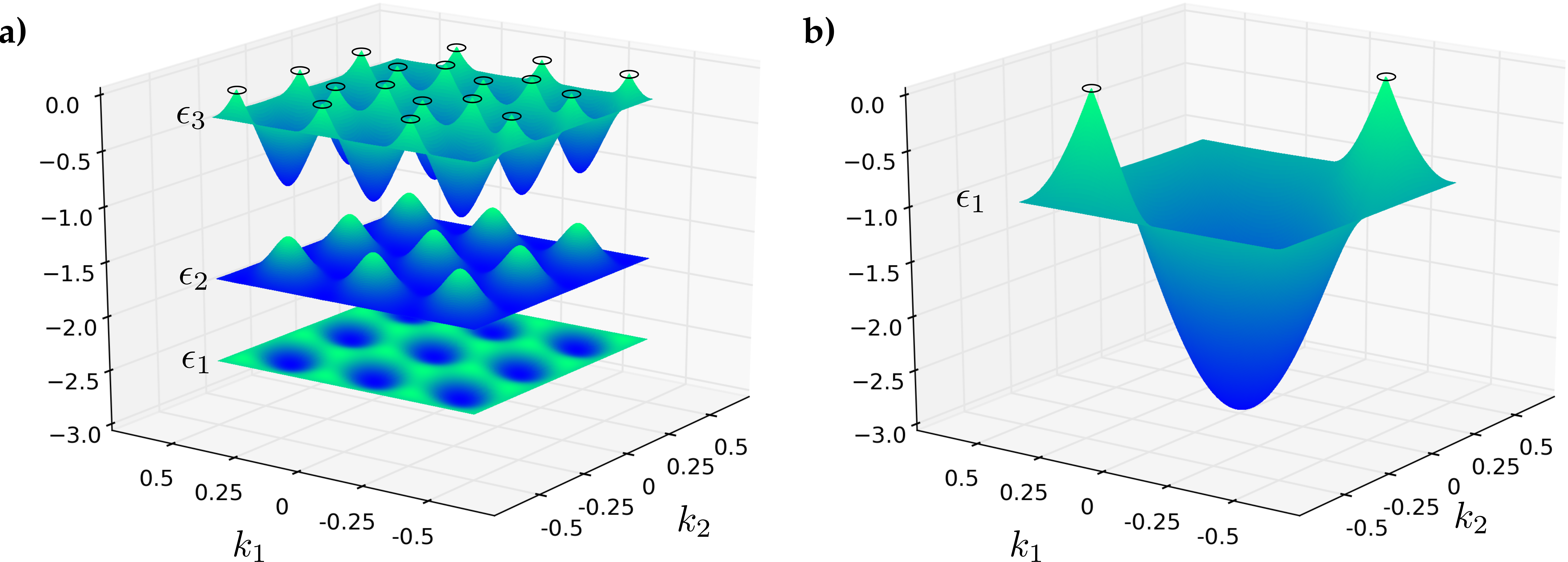}
	\caption{\label{fig:bulkfullk}  
	(a) Bulk spectrum (in units of the tunneling rate $t$) of the Bloch Hamiltonian 
	$\mathcal{H}_{\mathbf{k}}$ given by  Eq.~\eqref{Hamil} in the first Brillouin zone $BZ$ ($|k_{1,2}| \leq 1/2$). 
	Since the full spectrum is symmetric with respect to the zero-energy plane, 
	we only plot the 3 negative bands $\epsilon_1 \leq \epsilon_2 \leq \epsilon_3 \leq 0$. 
	The small black circles point out the location of Dirac points occurring between the energy bands 
	$\epsilon_3$ and $\epsilon_4$.
	 (b) Band structure of graphene. 
	 The small black circles point out the location of the 2 Dirac points of graphene.
	 Without the spin-orbit coupling term, our system is 
	 equivalent to graphene with threefold degenerate bands (i.e. one copy of the graphene 
	 band structure per spin component). 
The spin-orbit terms couple these 3 bands, 
eventually lifting their degeneracy and leading to the spectrum shown in (a). 
Due to the particular choice of the spin-orbit coupling terms $\mathcal{U}$ and $\mathcal{D}$, 
the band structure has the additional translation symmetry: 
$\epsilon_n(k_1,k_2)=\epsilon_n(k_1+1/3,k_2)=\epsilon_n(k_1,k_2+1/3)$.}
\end{figure*}

Diagonalizing the Bloch Hamiltonian $\mathcal{H}_{\mathbf{k}}$ 
yields 6 bands $\epsilon_1(\mathbf{k}) \leq \hdots \leq \epsilon_6(\mathbf{k})$ 
where the 3 (negative) lower bands are mirror images of the 3 (positive) upper bands with 
respect to the zero-energy plane, see Fig.~\ref{fig:bulkfullk}. This mirror symmetry originates 
from the bipartite nature of the honeycomb lattice and is also found in the usual band structure 
of graphene: eigenvalues come in opposite pairs. This is because 
$P\mathcal{H}_\mathbf{k}P =  -\mathcal{H}_\mathbf{k}$ where $P$ is the diagonal matrix 
with entries $\openone$ and $-\openone$. Noting that
\begin{align}
\mathcal{H}^2_{\mathbf{k}}= t^2 \left(
					\begin{matrix}
						N_{\mathbf{k}}^{\dagger}N_{\mathbf{k}} & 0\\
						0 &  N_{\mathbf{k}}N_{\mathbf{k}}^{\dagger}
					\end{matrix}
				\right),
\end{align}
where $N_{\mathbf{k}} = \exp(-\imag \mathbf{k}\cdot \bm{\delta}_1) \, M_{\mathbf{k}}$,
it is easy to show that $\epsilon^2_n(\mathbf{k}) = \mu^2_a(\mathbf{k}) \, t^2$ ($a=1,2,3$) 
where $\mu_3^2(\mathbf{k}) \leq \mu_2^2(\mathbf{k}) \leq \mu_1^2(\mathbf{k})$ 
are the eigenvalues of $N_{\mathbf{k}}^{\dagger}N_{\mathbf{k}}$ (and also of $N_{\mathbf{k}}N_{\mathbf{k}}^{\dagger}$).
We thus get the band structure $\epsilon_{1,2,3}= - t \, \mu_{1,2,3}$ and $\epsilon_{4,5,6} = t \, \mu_{3,2,1}$, 
highlighting the mirror symmetry of the bands with respect to the zero-energy plane. 

Dirac points are found at points $\mathbf{k} \in BZ$ where two eigenvalues of 
$N_{\mathbf{k}}^{\dagger}N_{\mathbf{k}}$ coalesce \cite{Comment2}. 
Noting that $N_{\mathbf{k}}^{\dagger} = S N_{\mathbf{-k}} S$, where $S$ is the anti-diagonal matrix with unit entries, 
we see that $N_{\mathbf{k}}^{\dagger} N_{\mathbf{k}}$ and $N_{\mathbf{-k}}N_{\mathbf{-k}}^{\dagger}$, 
and thus $N_{\mathbf{-k}}^{\dagger} N_{\mathbf{-k}}$, have the same spectrum. 
This shows that the Dirac points must come in opposite pairs in the Brillouin zone. 
In contrast to the graphene band structure which exhibits only 1 pair of such Dirac points, 
we get 9 pairs of Dirac points here, obtained when $\mu_3(\mathbf{k})=0$, 
that is between bands $\epsilon_3(\mathbf{k})$ and $\epsilon_4(\mathbf{k})$. 
The other bands remain fully isolated. Hence, just like graphene, our system is a semi-metal. 
We note that getting 9 pairs of Dirac points is not generic but specific to our $SU(3)$ model. 
It arises from an additional symmetry in our Hamiltonian due to our particular choice of the 
hopping matrices $\mathcal{U}$ and $\mathcal{D}$. 
We point out that this is not because of a smaller effective Brillouin zone since our Hamiltonian, 
as can be seen from the expression of $M_{\mathbf{k}}$ above, is invariant under Bravais translations only. 
Instead, one can show that 
\begin{align}
	N_{\mathbf{k}+\mathbf{b}_1/3}=\mathcal{U} N_{\mathbf{k}}\mathcal{U}^\dagger \\
	N_{\mathbf{k}+\mathbf{b}_2/3}=\mathcal{D}^{\dagger}N_{\mathbf{k}}\mathcal{D}, 
\end{align}
meaning that the matrices $N_{\mathbf{k}+\mathbf{b}_i/3}$ ($i=1,2$) and $N_{\mathbf{k}}$ are unitarily equivalent. 
Thus, up to a global gauge transform, 
the two Hamiltonians $\mathcal{H}_{\mathbf{k}+\mathbf{b}_1/3}$ and $\mathcal{H}_{\mathbf{k}}$ are identical. 
The same conclusion holds true for Hamiltonians $\mathcal{H}_{\mathbf{k}+\mathbf{b}_2/3}$ and $\mathcal{H}_{\mathbf{k}}$. 
Going back to the direct lattice, these global gauge transforms amount to a "rotation" of the spin 
degrees of freedom on each individual lattice site by the same unitary matrix. 
This shows that the Hamiltonians $\mathcal{H}_{\mathbf{k}+\mathbf{b}_1/3}$, 
$\mathcal{H}_{\mathbf{k}+\mathbf{b}_2/3}$ and $\mathcal{H}_{\mathbf{k}}$ have the same spectra. 
We thus infer that $\epsilon_n(\mathbf{k}) = \epsilon_n(\mathbf{k} +\mathbf{b}_1/3) = 
\epsilon_n(\mathbf{k} +\mathbf{b}_2/3)$ for each energy band.

Note that a mass term such as $\Delta(\hat{n}_A-\hat{n}_B)/2$ does not break the mirror 
symmetry between positive and negative eigenvalues, nor does it break the translation invariance 
of the lattice Hamiltonian. But it does lift the band degeneracies. Indeed, it adds the 
terms $\Delta /2$ and $-\Delta /2$ on the diagonal entries of $\mathcal{H}_{\mathbf{k}}$, 
Eq.\eqref{Hamil}, and the constant term $\Delta^2/4$ on the diagonal entries of 
$\mathcal{H}^2_{\mathbf{k}}$, leading to 
$\epsilon_n(\mathbf{k}, \Delta) =\pm t \, \sqrt{\mu_a^2(\mathbf{k})+(\Delta/2t)^2}$ and 
thus to $\epsilon_n(\mathbf{k},\Delta) = -\epsilon_{7-n}(\mathbf{k},\Delta)$. 
This shows that $C_n(\Delta) = C_{7-n}(\Delta)$: the Chern numbers come in equal 
pairs \cite{Comment3}. Because of this symmetry, the total zero-sum rule of 
Chern numbers boils down to 2 separate zero-sum rules $C_1+C_2+C_3=C_4+C_5+C_6=0$.

We now turn to the numerical computation of the Chern numbers. Strictly speaking, $C_n$, 
as given by Eq.~\eqref{eq:chern}, is only well-defined for an isolated band. Since the 2 
middle bands $\epsilon_3$ and $\epsilon_4$ are touching, we add a small mass term $\Delta=0.1t$, such that the six bands 
are fully isolated. Following~\cite{Fukui2005}, this allows to safely compute $C_n$ 
for each band and quantify the topology of the band structure. 
For our $SU(3)$ model, we find $C_1=3$, $C_2=-6$ and $C_3=3$~\cite{Comments4}. 
These nonzero values signal non-trivial topological properties of the bulk band structure.

\subsection{Finite system - Edge states}

\begin{figure}[t!p]
	\includegraphics[width=\linewidth]{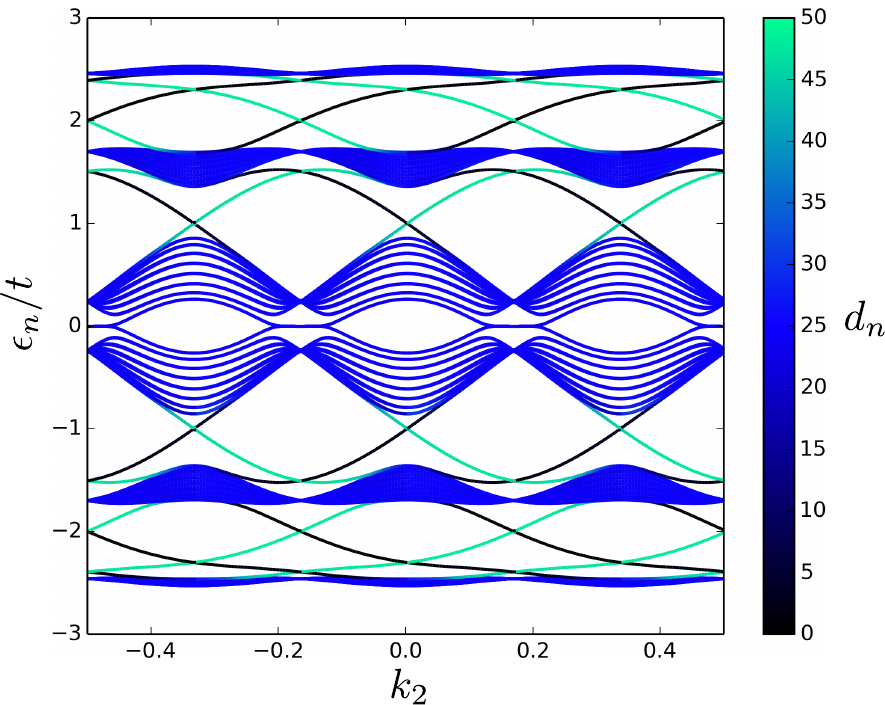}
	\caption{\label{fig:specloczoom} Spectrum, in units of the tunneling rate $t$, obtained by diagonalizing the 
	Hamiltonian Eq.~\eqref{eq:blochk2} describing a finite lattice strip with $N=50$ cells along $\mathbf{a}_1$ 
	(open boundary conditions) and periodic boundary conditions along $\mathbf{a}_2$. The mass term is $\Delta=0$.
	 Compared to Fig.~\ref{fig:bulkk2}, we see the appearance of topological 
	edge states in the gaps between the bulk bands with different Chern numbers. 
	No such edge states develop between bulk bands $\epsilon_3$ and $\epsilon_4$ 
	since they have equal Chern numbers. The color scale on the right side shows the average 
	position $d_n$ of the eigenstates along $\mathbf{a}_1$. As one can see the topological states connecting bands live 
	on either sides of the strip while the other bulk-like states fill the whole strip.
	}
	\end{figure}

\begin{figure}[t!p]
	\includegraphics[width=\linewidth]{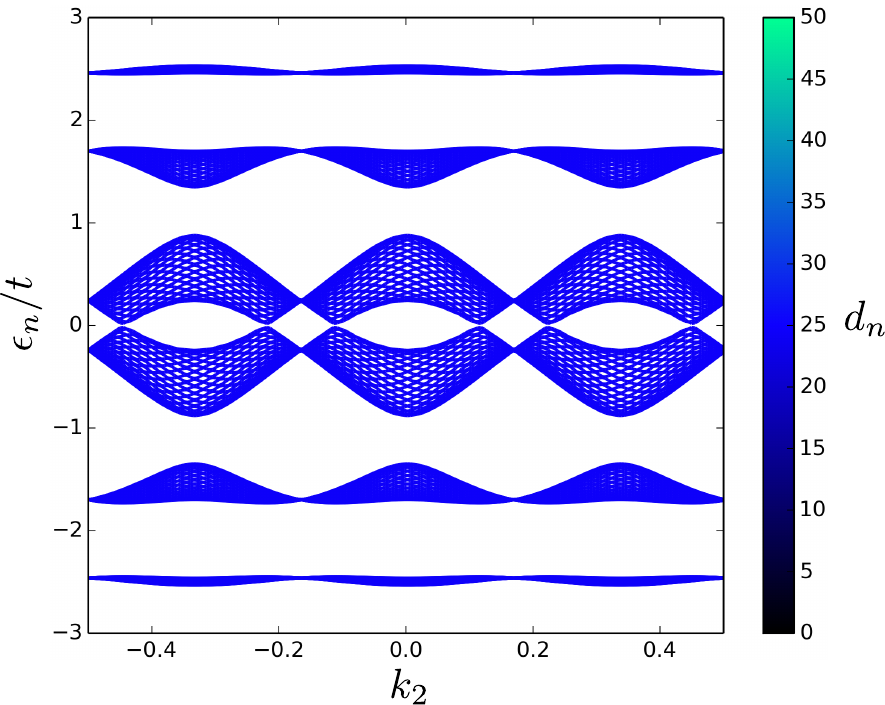}
	\caption{\label{fig:bulkk2} Bulk spectrum, in units of the tunneling rate $t$, 
	obtained by diagonalization of Hamiltonian Eq.~\eqref{Hamil} with periodic boundary conditions 
	both along $\mathbf{a}_1$ and $\mathbf{a}_2$. The number of cells along 
	$\mathbf{a}_1$ is $N=50$ and the mass term is $\Delta=0$. 	}
\end{figure}

As is well known, a lattice Hamiltonian having a band structure with non-vanishing Chern numbers 
exhibits edge states in a strip geometry~\cite{Kane2010}.
We consider such a finite lattice strip for our $SU(3)$ model and 
compute the spectrum for open boundary conditions in the $\mathbf{a}_1$ direction ($0 \leq n_1 \leq N$) 
and periodic boundary conditions in the $\mathbf{a}_2$ direction (unrestricted $n_2$). 
The lattice strip we consider has left and right zig-zag boundaries. 
Since the Hamiltonian is still invariant under lattice translations along $\mathbf{a}_2$, 
the Bloch wavevector along $\mathbf{b}_2$ remains a good quantum number. 
Therefore, we introduce the Fourier transform operators $\hat{\mathbf{a}}_{n_1,k_2}$ and 
$\hat{\mathbf{b}}_{n_1,k_2}$ along $\mathbf{a}_2$.
The Hamiltonian, for a given Bloch wavevector $k_2$, now reads:
\begin{align}
\label{eq:blochk2}
\hat{H}_{k_2}=-t\sum_{n_1=0}^N& \hat{\mathbf{b}}^{\dagger}_{n_1,k_2}
	(e^{i\mathbf{k}_2\cdot\bm{\delta}_1}\openone
	+e^{i\mathbf{k}_2\cdot\bm{\delta}_3} \mathcal{U}) \, \hat{\mathbf{a}}^{\phantom{\dagger}}_{n_1,k_2}
	\nonumber\\
	-t\sum_{n_1=1}^N& \hat{\mathbf{b}}^{\dagger}_{n_1-1,k_2}e^{i\mathbf{k}_2\cdot\bm{\delta}_2}\mathcal{D}
	\, \hat{\mathbf{a}}^{\phantom{\dagger}}_{n_1,k_2}+\mathrm{H.c.}
\end{align}
with $|k_2| \leq 1/2$.
By diagonalizing $\hat{H}_{k_2}$, we obtain the band spectrum shown in Fig.~\ref{fig:specloczoom}. 
This spectrum has to be compared to the bulk spectrum, shown in Fig.~\ref{fig:bulkk2}, 
that has been obtained for the same number of cells $N$ along $\mathbf{a}_1$ but with periodic boundary conditions. As expected from our analysis of the bulk spectrum, we indeed find additional states in the gaps between bands with different 
Chern numbers, that is between the first and second bands, between the second and third bands and, by
symmetry, between the fourth and fifth bands and between the fifth and sixth bands. 
No additional states appear between the third and the fourth band, since these two bands have equal Chern numbers. 

Let us define the position $x_i$ ($i=1, \hdots, 2N$) of the sites along a given 
horizontal zig-zag chain crossing the lattice strip from left to right. We denote by $P_n(x_i, k_2)$ 
the spatial distribution of the eigenstate with eigenvalue $\epsilon_n(k_2)$ along this chain. 
The color scale of Fig.~\ref{fig:specloczoom} shows the average position 
$d_n = \sum_i x_i P_n(x_i, k_2)$ of the eigenstate in the strip. We see that states chosen 
within the bulk-like bands spread uniformly over the whole lattice strip and $d_n$ lies at the center 
of the lattice strip. On the contrary, states connecting bulk-like bands with different Chern numbers 
are either strongly localized on the left boundary (black) or on the right boundary (green) of the lattice: 
these are the celebrated edge states. 
We also see that, in our system, edge states within a given gap are localized on one side 
of the strip when their group velocity is positive and on the other side of the strip when 
their group velocity is negative. This one-to-one correspondence between the sign of the slope of 
the energy dispersion relation of an edge state and its spatial localization changes from one gap to the other. 
This feature, particular to our system, emphasizes the chiral character expected for particle transport at 
the boundaries. Interestingly, one can recover the values of the Chern numbers of our bulk system from 
the bulk-edge correspondence ~\cite{Hatsugai1993a,Hatsugai1993b,Hatsugai1997}. Within a given gap, 
one counts the number of edge state $N_+$ and $N_-$ with positive and negative group velocities 
that give rise to a localization on the right side of the strip. The difference $(N_+-N_-)$ is 
then equal to the sum of the Chern numbers of all the bands below the gap considered. This recipe 
allows to reconstruct all Chern numbers with their sign. It is easy to check that we do recover 
the values computed in the preceding paragraph. 
For instance, the absence of edge states linking the third and the fourth bands ($N_+=N_-=0$) is 
in agreement with the fact that the sum of the Chern numbers of the three lowest bands vanishes, $N_+-N_- = C_1+C_2+C_3=0$.

\begin{figure}[t!p]

	\centerline{\includegraphics[width=0.45\textwidth]{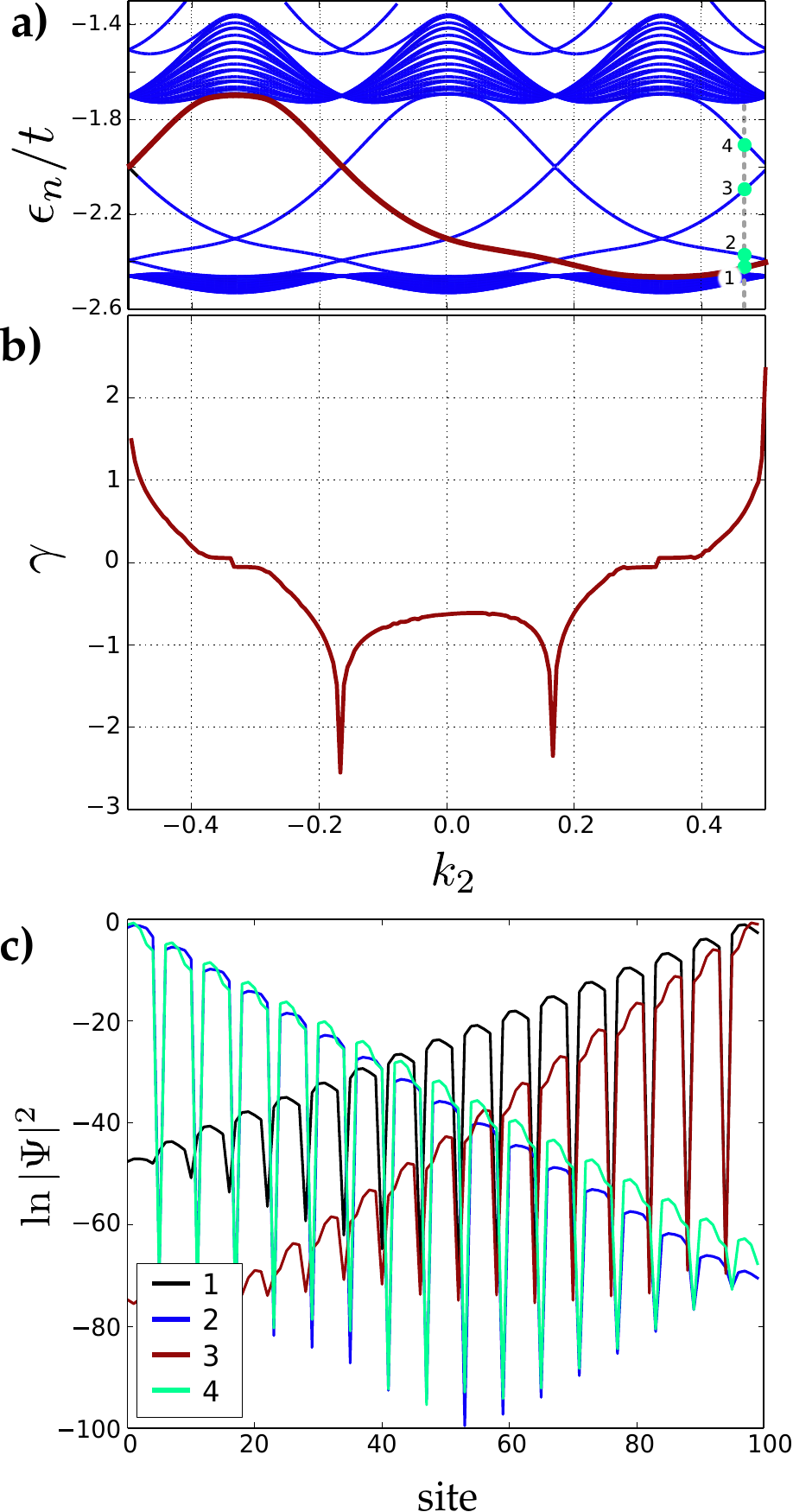}}
	\caption{\label{fig:edgelock} Localization properties of edge states. 
	(a) Zoom of the band structure shown in Fig.\ref{fig:specloczoom}. 
We consider the edge states associated to the eigenvalues highlighted in red and 
the edge states 1, 2, 3 and 4 obtained at the particular value $k_2\simeq0.458$. 
Edge state 1 and 3 have positive group velocities, while states 2 and 4 have negative group velocities. 
The probability distribution $P_n(n_1, k_2)$ of these states behave 
like $\exp(\gamma n_1)$ with $\gamma \geq 0$ when the state is localized 
on the right side of the lattice strip and $\gamma \leq 0$ when it is localized on its left side. 
(b) Plot of $\gamma(k_2)$ for the edge states with energy dispersion relation highlighted in 
red in panel (a). As one can see, states with negative group velocities are localized 
at $n_1=0$ and $\gamma < 0$. For values of $k_2$ roughly between $0.6$ and $0.7$, the states 
dive into the bulk band, become delocalized and $\gamma$ vanishes. 
For larger $k_2$ values, the group velocity becomes positive, the edge states are now
localized at $n_2=N$ and $\gamma >0$. (
c) Plot of $\ln P_n$ as a function of $x_i$ along a horizontal zig-zag chain crossing 
the lattice strip for the edge states 1, 2, 3 and 4 indicated in panel (a). 
Their localization is correlated with the sign of their group velocity. 
We see that the closer the state to the bulk-like band, the smaller $\gamma$. 
}
\end{figure}

The probability distribution $P_n(x_i, k_2) \propto e^{\gamma x_i}$ of an edge state decays exponentially 
with $x_i$ when it is localized at $n_1=0$ ($\gamma <0$) and grows exponentially with $x_i$ 
when it is localized at $n_1=N$ ($\gamma >0$). The (positive or negative) 
value of the characteristic scale $1/\gamma$ 
depends both on the edge state considered and on the Bloch wavector $k_2$. 
This is emphasized in Fig.~\ref{fig:edgelock}b where we plot $\gamma$ as a function of $k_2$ for 
the edge states with the dispersion relation highlighted in Fig.~\ref{fig:edgelock}a. As one can see,  the state is
localized on the left side of the strip ($n_1=0$) as long as its group velocity is negative. 
For values of $k_2$ roughly in the range $0.6 - 0.7$, the state dives into the
bulk band and becomes delocalized. The decay coefficient $\gamma$ then vanishes. For larger $k_2$ values,
the group velocity becomes positive and the state is now localized on the right 
side of the strip ($n_1=N$). Finally,  Fig.~\ref{fig:edgelock}c shows 
the probability distribution $P_n$ for the different edge states shown in 
Fig.~\ref{fig:edgelock}a) for a given value of $k_2\simeq0.458$. Here again, one can see the 
correlation observed in our system between the sign of the group velocity and the localization center of the edge state.

\section{Topological transitions}
\label{sec:transition}

The Chern number of a given isolated band is an invariant integer-valued topological quantity. 
Its value can only change when the band comes in contact with another one. A vanishing gap can 
be achieved by changing parameters in the Hamiltonian, for example the strength of the spin-orbit 
coupling. In general, the hopping matrices along links $\bm{\delta}_a$ can be written as 
$T_a = -t\exp{(\imag A_a)}$ ($a=1,2,3$). The traceless Hermitian featuring gauge field can
be written as $A_a=\bm{\alpha}_a\cdot\bm{\lambda}$, where $\bm{\alpha}_a$ is a real 
8-component vector and $\bm{\lambda}$ is the 8-component vector made of the Gell-Mann matrices. 
We have first monitored the band structure, gaps and Chern numbers by considering various 
configurations of the three vectors $\bm{\alpha}_a$. However, for all configurations that we 
tested, we could only recover the set of Chern numbers $(3,-6,3)$ already found in 
Section~\ref{sec:topo} for the 3 negative bands, or the opposite set $(-3,6,-3)$. 
The same result was obtained by allowing for imbalanced tunneling amplitudes, that is 
$T_a=-t_a\exp{(\imag A_a)}$, and by varying the $t_a$. 

More saliently, we found other topological transitions giving rise to different sets 
of Chern numbers by adding a spin-dependent chemical potential term $-(\mu_A n_A + \mu_B n_B)$ to the Hamiltonian. 
The chemical potential matrix for $A$ sites reads 
$\mu_A = -\Delta/2 \, \openone+\delta\mu_A S_z$ and 
$\mu_B = \Delta/2 \, \openone+\delta\mu_B S_z$ for $B$ sites. 
Note that $\Delta$ is the usual (spin-independent) mass term (see Section~\ref{sec:topo}).  
All in all, we found that the two following spin-dependent  chemical potential configurations
        \begin{itemize}
         \item[$\bullet$] $\delta\mu_B=-\delta\mu_A=\Delta_S/2$ (spin-dependent mass imbalance).
                        
         \item[$\bullet$] $\delta\mu_B=\delta\mu_A$ (spin-dependent local potential). 
         
        \end{itemize} 
with an additional spin-orbit coupling along $\bm{\delta}_1$ gave rise to a 
larger variety of Chern numbers. We point out that, without this additional spin-orbit coupling, 
one can still find Chern numbers differing from $\pm(3,-6,3)$, but only for some particular 
values of the spin-orbit couplings along $\bm{\delta}_2$ and $\bm{\delta}_3$.
        
\begin{table}[t!p]
\begin{ruledtabular}
\begin{tabular}{c|r|r|r|r|r|r|r|r|r|r|r|r}
$C_6$ & 0 &  0 &  0 & -1 & -2 & -2 & -3 & -3 & -2 & -1 &  0 & 0 \\ 
\hline
$C_5$ & 0 &  1 &  2 &  3 & 4 &  5 &  6 &  5 &  4 &  3 &  1 & 0 \\
\hline
$C_4$ & 0 & -1 & -2 & -2 & -2 & -3 & -3 & -2 & -2 & -2 & -1 & 0 
\end{tabular}
\end{ruledtabular}
\caption{\label{tab:chernmu} The different sets of Chern numbers obtained for the 3 
highest bands for increasing values of $\Delta_S$, see Fig.~\ref{fig:chernmu}. At 
$\Delta_S=0$, one recovers the expected set of Chern numbers $(-3,6,-3)$. 
The values for the 3 lowest bands are obtained from the symmetry relation 
$C_{7-n}(\Delta_S) = C_n(-\Delta_S)$. They satisfy the zero-sum rule $C_6+C_5+C_4=0$ 
for each value of $\Delta_S$, see text.} 
\end{table}

\begin{figure*}[t!p]
\centerline{\includegraphics[width=0.9\linewidth]{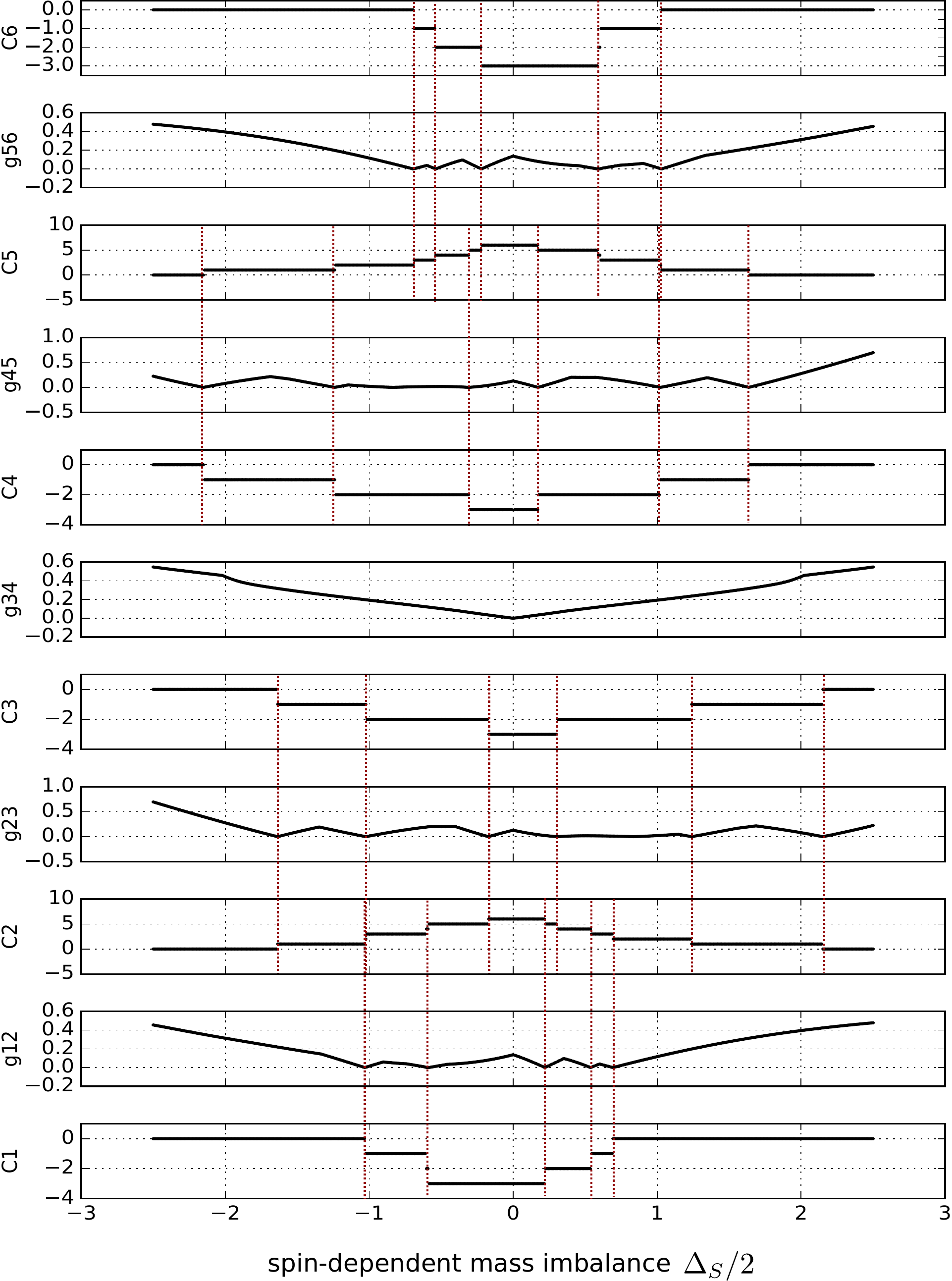}}
\caption{\label{fig:chernmu} Chern numbers $C_n$ and band 
gaps $g_{n+1,n} = \mathrm{Min} \left[\epsilon_{n+1}(\mathbf{k}, \Delta_S)-\epsilon_n(\mathbf{k}, \Delta_S)\right]$ as
functions of the spin-dependent mass imbalance $\Delta_S$. Many topological transitions are
observed, resulting in various integer changes of the Chern number values, 
which are no longer restricted to $\pm(3,-6,3)$ (see text). The red-dotted vertical lines 
are guides to the eye emphasizing that
a change in Chern numbers is always associated with a vanishing gap between two bands. 
Due to a symmetry of the Hamiltonian, we have $C_{7-n}(\Delta_S)=C_{n}(-\Delta_S)$, see text. 
For $\Delta_S=0$, one recovers the expected set of Chern numbers $(-3,6,-3)$. }
\end{figure*}

\begin{figure}[t!p]
 
 \centerline{\includegraphics[width=\linewidth]{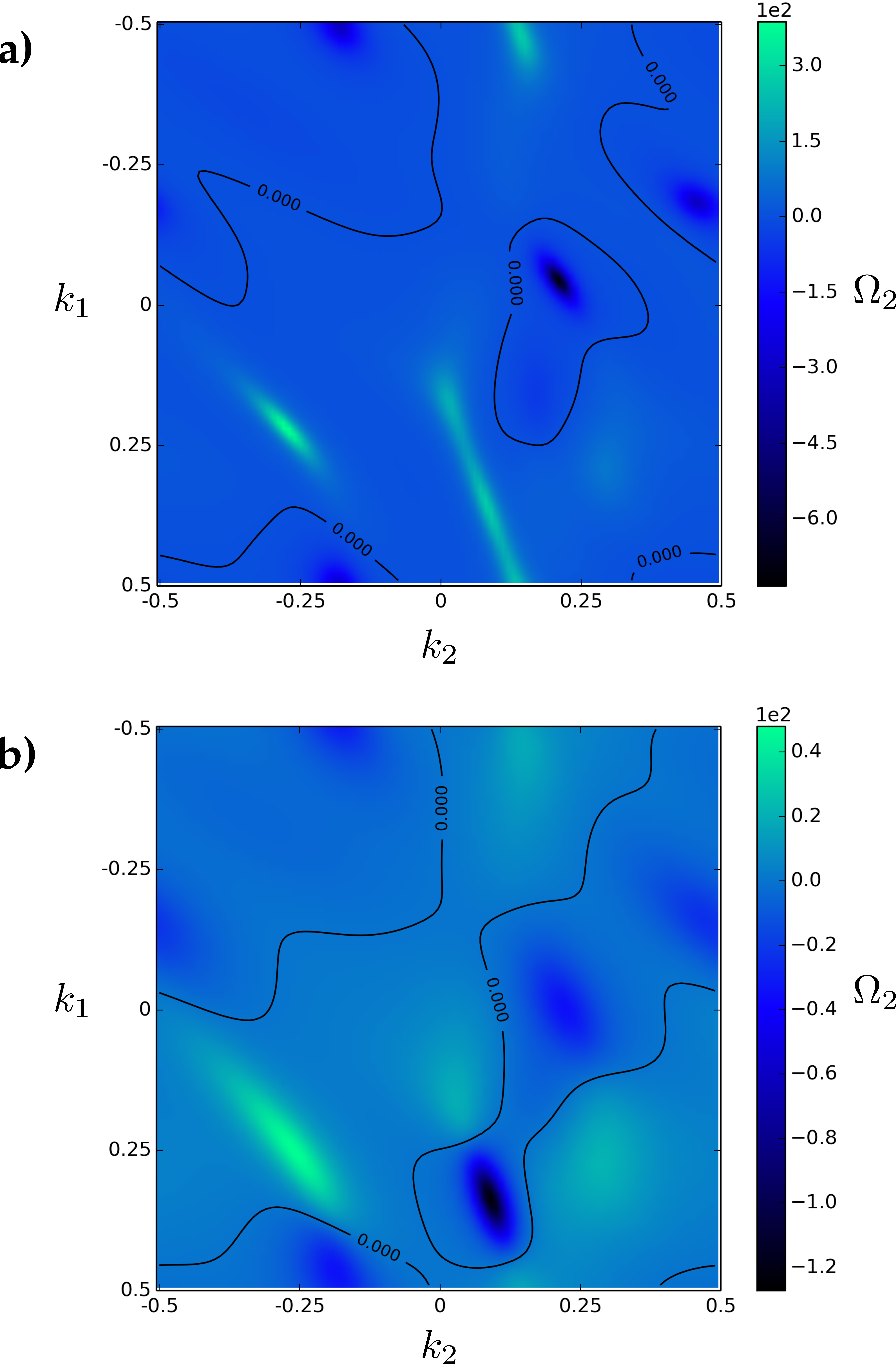}}

\caption{\label{fig:berrycurvature} Plot of Berry curvature $\Omega_2(\mathbf{k}, \Delta_S)$ 
and the level curve $\Omega_2=0$ for the second band $\epsilon_2(\mathbf{k}, \Delta_S)$, 
see Fig.~\ref{fig:chernmu}. Top plot: $\Delta_S=-2.6$ ($C_2 = 1$). Bottom plot: $\Delta_S=-4$ ($C_2=0$). 
Both Berry curvatures exhibit qualitatively similar structures. 
In particular, there is no obvious particular pattern or symmetry explaining why $C_2$ 
vanishes when $\Delta_S=-4$. }
\end{figure}

As an example, we consider a system with spin-dependent mass imbalance ($\Delta_S \neq 0$) 
and non-Abelian hopping matrices $T_1=-t \,\mathcal{V}=-t \exp{(\imag A_1)}$ where 
$A_1=\frac{2\pi}{3\sqrt{3}}(\lambda_1+\lambda_4+\lambda_6)$, $T_2 = - t \, \mathcal{D}$ 
and $T_3 = -t \, \mathcal{U}$. For sake of simplicity, we also choose $\Delta =0$. 
Fig.~\ref{fig:chernmu} shows the 6 Chern numbers $C_n$ and the different band gaps 
that are obtained as a function of $\Delta_S$. As one can see, we get new sets 
of Chern numbers now and not just $\pm(3,-6,3)$, see Table~\ref{tab:chernmu}. 
The vertical lines in Fig.~\ref{fig:chernmu} are guides to the eye that mark a gap 
closing between 2 adjacent bands. We see that the different gaps do not close at the same time. 
This emphasizes that it is only the pair of Chern numbers $C_{n+1}$ and $C_n$ of the 
bands involved in the gap closing $g_{n+1,n}=0$ that can change. One has: 
\begin{align}
\left(C_{n+1}+C_n\right)_{\textrm{after}} = \left(C_{n+1}+C_n\right)_{\textrm{before}},
\end{align}
the other Chern numbers remain unaffected. We further note that changing the sign 
of all the mass imbalances amounts to flipping the sign of the eigenvalues of the 
Hamiltonian, $\epsilon_n(\mathbf{k}, -\Delta,-\Delta_S)= - \, \epsilon_{7-n}(\mathbf{k}, \Delta,\Delta_S)$. 
Again, this is because $P\mathcal{H}_\mathbf{k}(\Delta, \Delta_S)P = - \mathcal{H}_\mathbf{k}(-\Delta, -\Delta_S)$.
We therefore have the property that $C_{7-n}(\Delta, \Delta_S)=C_{n}(-\Delta,-\Delta_S)$. 
This symmetry in the Chern numbers can be readily checked in Fig.~\ref{fig:chernmu} obtained for $\Delta=0$. 
One also confirms that gaps have the same symmetry, $g_{n+1,n}(\Delta_S) = g_{7-n,6-n}(-\Delta_S)$, 
from which one concludes that $g_{43}$ is symmetric in $\Delta_S$. 
We also note that the gap $g_{43}$ only closes at $\Delta_S=0$ but that this degeneracy 
does not modify $C_3$ and $C_4$. As a consequence, the change in Chern numbers after a 
gap closing can only occur within the subset $(C_1, C_2, C_3)$ or within $(C_4,C_5,C_6)$. 
The Chern numbers thus satisfy again the separate zero-sum rules $C_1+C_2+C_3 = C_4+C_5+C_6 = 0$.
Finally all bands eventually reach a vanishing Chern number for large enough 
values of $|\Delta_S|$, even though the Berry curvatures neither vanish nor 
display any particular symmetry. Fig.~\ref{fig:berrycurvature} shows a plot 
of the Berry curvature of the second band $\epsilon_2(\mathbf{k}, \Delta_S)$ 
in the Brillouin zone obtained for $\Delta_S = -2.6$ where $C_2=1$ (top plot) 
and for $\Delta_S=-4$ where $C_2=0$ (bottom plot). In both cases, the Berry 
curvatures exhibit qualitatively similar structures. 

As explained above, such a rich variety of Chern numbers and topological 
transitions have been obtained by imposing spin-dependent chemical potentials 
on sites $A$ and $B$ \textit{and} different  spin-orbit couplings along the 3 links, 
keeping intact, at the same time, the underlying triangular Bravais symmetry of 
the honeycomb lattice. We have studied the same situation on the square lattice 
(not shown here) but were not able to find such a rich variety of Chern numbers 
and topological transitions. The probable reason is that the square lattice 
has only 2 link vectors and thus accommodates less configurations of spin-orbit couplings.  
From that point of view, the $SU(3)$ model on the honeycomb lattice exhibits a 
band structure with richer topological properties. 
        
A similarly rich variety of topological transitions and Chern numbers is obtained 
by allowing for complex $\bm{\alpha}_a$ in the gauge fields 
$A_a= \bm{\alpha}_a \cdot \bm{\lambda}$. However, this situation, even if 
experimentally feasible, does not correspond to a pure $SU(3)$ gauge potential anymore.  
        
\begin{figure}[t!p]
	\includegraphics[width=\linewidth]{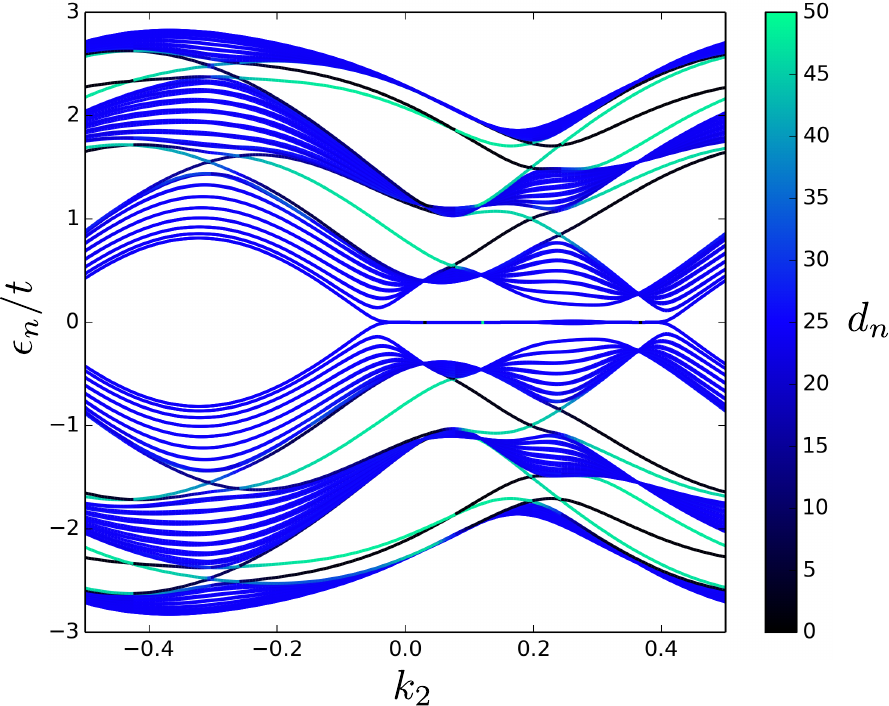}
	\caption{\label{fig:nonisbands} Band structure (in units of the tunneling rate $t$) 
	for the strip lattice configuration when 
	$\Delta=\Delta_S=0$, $T_1=-t\mathcal{U}$, $T_2=-t\mathcal{D}$ and $T_3=-t\mathcal{V}$ (see text).
	Even though the bulk-like bands are well isolated and thus never touch anywhere in the Brillouin zone,  
there is no finite \textit{charge} gap separating them anymore. 
The Chern numbers being well defined, edge  states are still present and link the different bands. 
The color scale on the right side shows the average position $d_n$ of the 
eigenstates along $\mathbf{a}_1$. Because of the absence of a charge gap, 
the system is not a Chern insulator. }
\end{figure}

We emphasize that the bulk bands remain well isolated for a large variety of 
gauge fields $A_a$ and for many values of the the spin-dependent mass imbalance $\Delta_S$. 
However, even though the bands do not touch at any point in the Brillouin zone, 
they can still overlap on the energy axis, which means that there is no  \textit{charge} gap 
separating them. This situation is depicted in Fig.~\ref{fig:nonisbands} for 
$\Delta=\Delta_S=0$, $T_1=-t \,\mathcal{U}$, $T_2=-t \, \mathcal{D}$ and 
$T_3=-t \, \mathcal{V}$. But, since the Chern numbers are still well defined, 
topological edge states linking the different bands still appear in the strip 
lattice configuration. On the other hand, the absence of charge gaps prevents 
the system from behaving like a Chern insulator and no Hall plateaus are then 
expected in transverse conductance measurements.

\section{Conclusion}
\label{sec:conclusion}
We have studied the topological properties
of a non-interacting Hubbard model on the honeycomb lattice with $SU(3)$ spin-orbit couplings. 
We have emphasized that, in marked contrast with lattice $SU(2)$ models which are 
always topologically trivial, these $SU(3)$ models can break time reversal invariance. 
As a consequence, their bulk band structure becomes topologically non trivial: 
bulk bands have non-zero Chern numbers and chiral edge states develop in a lattice strip configuration with open boundary conditions. 
We have also shown that $SU(3)$ models on the honeycomb lattice allow for a larger variety of topological transitions than on the square lattice~\cite{Barnett2012} because of a higher lattice coordination number. 

A natural extension of the present work is to include interactions in our model and understand their impact on the topological properties of the system, in particular the emergence of non-trivial spin textures
breaking the translation symmetry of the lattice~\cite{Cole_2012,Hofstetter12,Grass2014,Mandal2016}. Because $SU(3)$ is a larger gauge group than $SU(2)$, we expect a much larger variety of topological properties 
for these spin textures~\cite{Hueda12,Ezawa_book}.

\appendix
	
\section{Gell-Mann matrices}
\label{app:GMM}
The generators of the $SU(3)$ group are $g_a = \lambda_a/2$ where the $\lambda_a$ are the Gell-Mann matrices:
\begin{equation}
\begin{aligned}
\lambda_{1}&={\begin{pmatrix}0&1&0\\1&0&0\\0&0&0\end{pmatrix}} \qquad
\lambda_{2}={\begin{pmatrix}0&-i&0\\i&0&0\\0&0&0\end{pmatrix}} \\
\lambda_{3}&={\begin{pmatrix}1&0&0\\0&-1&0\\0&0&0\end{pmatrix}} \qquad
\lambda_{4}={\begin{pmatrix}0&0&1\\0&0&0\\1&0&0\end{pmatrix}}  \\
\lambda_{5}&={\begin{pmatrix}0&0&-i\\0&0&0\\i&0&0\end{pmatrix}} \qquad	
\lambda_{6}={\begin{pmatrix}0&0&0\\0&0&1\\0&1&0\end{pmatrix}} \\
\lambda_{7}&={\begin{pmatrix}0&0&0\\0&0&-i\\0&i&0\end{pmatrix}}  \qquad	
\lambda_{8}={\frac {1}{\sqrt {3}}}{\begin{pmatrix}1&0&0\\0&1&0\\0&0&-2\end{pmatrix}}
\end{aligned}
\end{equation}

They satisfy Tr$(\lambda_a\lambda_b) = 2\delta_{ab}$. One can easily check that: 

\begin{equation}
\begin{aligned}
\Theta\lambda_1\Theta^{-1}&=-\lambda_6 \qquad \Theta\lambda_2\Theta^{-1}=-\lambda_7 
\qquad \Theta\lambda_4\Theta^{-1}=\phantom{-}\lambda_4 \\
\Theta\lambda_5\Theta^{-1}&=\phantom{-}\lambda_5 \qquad \Theta\lambda_6\Theta^{-1}=-\lambda_1 
\qquad \Theta\lambda_7\Theta^{-1}=-\lambda_2 \\
\Theta\lambda_3\Theta^{-1}&=\phantom{-}\frac{\lambda_3-\sqrt{3}\lambda_8}{2}\\
\Theta\lambda_8\Theta^{-1}&=-\frac{\lambda_8+\sqrt{3}\lambda_3}{2}.
\end{aligned}
\end{equation}

\section{Time-reversal and unitary matrices}
\label{app:TRU}

Time reversal symmetric unitary matrices $U \in U(3)$ can be parametrized as follows
\begin{widetext}
\begin{equation}
	U_{even}= \pm \frac{1}{1+\cos^2\phi}\left(\begin{matrix}
		2\cos^2\phi e^{i\chi_a} & 2\sin\phi\cos\phi e^{\frac{i}{2}(\chi_a+\chi_b)}e^{i\pi n} & \sin^2\phi e^{i\chi_b}\\
		-2\sin\phi\cos\phi e^{\frac{i}{2}(\chi_a-\chi_b)}e^{i\pi n} & 2-3\sin^2\phi & 2\sin\phi\cos\phi e^{-\frac{i}{2}(\chi_a-\chi_b)}e^{-i\pi n}\\
		\sin^2\phi e^{-i\chi_b} & -2\sin\phi\cos\phi e^{-\frac{i}{2}(\chi_a+\chi_b)}e^{-i\pi n} & 2\cos^2\phi e^{-i\chi_a}
	\end{matrix}\right), 
\end{equation}
\end{widetext}	
where the angles $\phi$, $\chi_a$ and $\chi_b$ are arbitrary real parameters and 
$n$ is an arbitrary integer. 
Since $\textrm{Det}(\Theta U_{even} \Theta^{-1}) = \textrm{Det}(U_{even}^*) = 
(\textrm{Det} \, U_{even})^* =  \textrm{Det} \, U_{even}$, 
we conclude that $\textrm{Det}\,U_{even}$ is real and $U_{even}$ is unimodular: 
$\textrm{Det}\,U_{even} =\pm 1$. This can be directly checked from the matrix expression given above. 
This result is in fact general: any unitary $N\times N$ matrix which is even under time reversal is unimodular. 

If $U$ is odd under time reversal, a convenient parametrization is: 
\begin{widetext}
\begin{equation}
	U_{odd}=\pm \frac{1}{1+\cos^2\phi}\left(\begin{matrix}
		2\cos^2\phi e^{i\chi_a} & 2\sin\phi\cos\phi e^{\frac{i}{2}(\chi_a+\chi_b)}e^{i\pi n} & \sin^2\phi e^{i\chi_b}\\
		-2i\sin\phi\cos\phi e^{\frac{i}{2}(\chi_a-\chi_b)}e^{i\pi n} & i(2-3\sin^2\phi) & 2i\sin\phi\cos\phi e^{-\frac{i}{2}(\chi_a-\chi_b)}e^{-i\pi n}\\
		-\sin^2\phi e^{-i\chi_b} & 2\sin\phi\cos\phi e^{-\frac{i}{2}(\chi_a+\chi_b)}e^{-i\pi n} & -2\cos^2\phi e^{-i\chi_a}
	\end{matrix}\right).
\end{equation}	
\end{widetext}
Now we have 
$(\textrm{Det} \, U_{odd})^* = \textrm{Det}(-U_{odd}) = -\textrm{Det} \, U_{odd}$ and $\textrm{Det} \, U_{odd} = \mp \imag$
is purely imaginary (as can be directly checked with the above matrix expression). 
As a consequence, $U_{odd}$ does not belong to $SU(3)$. 
Alternatively, $SU(3)$ matrices breaking time reversal invariance 
($\Theta \mathcal{U} \Theta^{-1} \neq \mathcal{U}$) cannot be odd. 
This results holds true for any unitary $N\times N$ matrix which is odd under 
time reversal provided $N$ is odd. 
Indeed $(\textrm{Det} \, U_{odd})^* = \textrm{Det}(-U_{odd}) = (-1)^N \textrm{Det} \, U_{odd}$ 
shows that $\textrm{Det} \,U_{odd}$ is purely imaginary when $N$ is odd. 
Thus $U_{odd}$ cannot belong to $SU(N)$ and $SU(N)$ matrices breaking time 
reversal invariance cannot be odd when $N$ is odd. When $N$ is even, $U_{odd}$ is unimodular. 
This means that it is possible to have $SU(N)$ matrices which are odd under time reversal when $N$ is even, 
with the notable exception of $N=2$, as shown in the paper.


\begin{thebibliography}{11}
\bibitem{Comment2} Triple degeneracies cannot occur because $N_{\mathbf{k}}^\dagger N_{\mathbf{k}}$ is never diagonal in the Brillouin zone, as can be checked by direct inspection.
\bibitem{Comment3} Noting that $P\mathcal{H}_\mathbf{k}(\Delta)P =  -\mathcal{H}_\mathbf{k}(-\Delta)$, we also infer that $C_n(\Delta) = C_n(-\Delta)$.
\end{thebibliography}
\end{document}